\documentclass[aps,showpacs,preprintnumbers,amsmath,amssymb]{revtex4}
\usepackage{dcolumn}
\usepackage[dvips]{epsfig}
\usepackage{graphicx}
\usepackage{amssymb}
\usepackage{color}
\usepackage{enumerate}

\begin{document}
\baselineskip=0.8 cm
\title{\bf Testing gravity of a disformal Kerr black hole in quadratic degenerate
higher-order scalar-tensor theories by quasi-periodic oscillations}

\author{Songbai Chen$^{1,2}$\footnote{Corresponding author: csb3752@hunnu.edu.cn},
Zejun Wang$^{1}$
Jiliang Jing$^{1,2}$ \footnote{jljing@hunnu.edu.cn}}
\affiliation{ $ ^1$ Department of Physics, Key Laboratory of Low Dimensional Quantum Structures
and Quantum Control of Ministry of Education, Synergetic Innovation Center for Quantum Effects and Applications, Hunan
Normal University,  Changsha, Hunan 410081, People's Republic of China
\\
$ ^2$Center for Gravitation and Cosmology, College of Physical Science and Technology, Yangzhou University, Yangzhou 225009, People's Republic of China}

\begin{abstract}
\baselineskip=0.6 cm
\begin{center}
{\bf Abstract}
\end{center}

By using the relativistic precession model, we have studied frequencies of quasi-periodic oscillations in the spacetime of a disformal Kerr black hole. This black hole owns an extra disformal parameter and belongs to a class of non-stealth solutions in quadratic degenerate higher-order scalar-tensor (DHOST) theories. Our result shows that only the periastron precession frequency is related to the disformal parameter, while the azimuthal frequency and the nodal precession frequency are identical with those in the usual Kerr black hole in general relativity. Combing with the observation data of GRO J1655-40, we fit parameters of the disformal Kerr black hole, and find that  the disformal parameter $\alpha$ is almost negative in the range of $1 \sigma$, which implies the negative disformal parameter $\alpha$ is favored by the observation data of GRO J1655-40.

\end{abstract}

 \maketitle
\newpage
\section{Introduction}

The observations of gravitational waves \cite{P1,P2,P3,P4,P5} together with the first image of the supermassive compact star M87* \cite{fbhs1,fbhs6} confirm the existence of black holes in nature. In Einstein's general
relativity,  the geometry of a neutral rotating black hole is described by so-called Kerr metric only with two parameters, i.e., the black hole mass and spin, which is consistent with the well-known no-hair theorem. Up to now, Einstein's theory of gravity has successfully passed a series of tests from observational experiments, and then it is natural to expect that an astrophysical
black hole in our Universe should be a Kerr black hole. However, there still exist a lot of research efforts on black hole solutions in alternative theories  because the observed accelerating expansion of the current Universe
still leaves ample room for such kinds of theories of gravity.

Scalar-tensor theories are a kind of very important alternative theories where one scalar degree of freedom is combined with the gravitational metric. At present, the most general family of scalar-tensor
theories is the so-called degenerate higher-order scalar-tensor (DHOST) theories \cite{dhost1,dhost2,dhost3,dhost4}, which contain the higher order derivatives of scalar field and meet a certain set of degeneracy conditions. Actually, DHOST theories can be also regarded as a kind of extensions of Horndeski theories \cite{late2}  and of beyond Horndeski  theories \cite{late3}. Especially, with the degeneracy conditions, DHOST theories are free from the Ostrogradsky ghost even if there exists higher-order equations of motion. This means that the degeneracy of Lagrangian \cite{dhost1,dhost2} is the crucial element characterized higher-order theories.
Generally, it is not an easy task to obtain exact black hole solutions in alternative theories of gravity due to the more complicated field equation. However, in the DHOST theories, some new black hole solutions have emerged recently \cite{c2633,c2634,c2635,dhost24,dhost26,dhost20,dhost27,dhost28}.  These solutions can be classified into two types: the stealth solutions and the non-stealth solutions. In the stealth solutions, there exists an extra scalar field, but the solution owns the same metric as that in general relativity because the scalar field is absented in the spacetime metric.  In the non-stealth solutions, the parameters of scalar field appear in the metric and then the metric form of solution differs from that in Einstein's theory.

It is well-known that the disformal and conformal transformations can take us from some DHOST Ia theory to some
other specific DHOST Ia theory \cite{dhost3,dhost4}. Thus, begin with a ``seed" known solution $\tilde{g}_{\mu\nu}$ in DHOST Ia
theory, one could obtain a new solution  $g_{\mu\nu}$ in other specific DHOST Ia theory by using this transformation \cite{dhost3,dhost4}.  Along this spirit, a disformal rotating black hole solution is obtained recently in quadratic DHOST theories \cite{dhostv1,dhostv2}. Besides usual mass and spin parameters, the disformal rotating black hole solution else owns an extra disformal parameter which describes the deviation from the Kerr geometry \cite{dhostv1,dhostv2}. The scalar hair carried by this solution is time-dependent with a constant kinetic density. As in the usual Kerr case, the disformed Kerr solution is asymptotically flat and has a single curvature singularity situated at $r^2+a^2\cos^2\theta =0$.  However, the presence of  the scalar field changes the spacetime geometry so that the disformed Kerr spacetime is no more Ricci flat. Moreover, the scalar field also modifies the position and width of ergoregions, and then changes the possibility of extracting rotation energy from black hole by the Penrose process. The recent study indicates \cite{shadow1} that the shadow of a disformal rotating black hole depends heavily on the deformation parameter. Especially, one can find that there are some eyebrowlike shadows and the self-similar fractal structures in the black hole shadow. These novel features originating from the scalar field
imply that the disformal rotating non-stealth black hole could give rise to some new observational effects differed from those in the Kerr case, which could provide some potential ways to constrain the deformation parameter and to examine further the corresponding quadratic DHOST theories in the strong field region.

Quasi-periodic oscillations are
seen as peaks in the X-ray power density spectrum emitted by accreting black hole binaries \cite{qpo1,qpo2}, which is widely believed as a useful tool to test theories of gravity in the strong field regime and to constrain black hole parameters.
There are several types of quasi-periodic oscillations in terms of their properties. In general, their low frequencies are distributed in the range $0.1\sim 30$ Hz and high-frequencies are in the range $100\sim500$ Hz. However, the exact mechanism for quasi-periodic oscillations is still lacking at present. The relativistic precession model has been proposed to explain twin high-frequency quasi-periodic oscillations as well as a low-frequency mode in low-mass X-ray binaries \cite{RPM1,RPM2,RPM20,RPM3,RPM301,RPM302}. In this precession model, quasi-periodic oscillation frequencies are assumed to associate with three fundamental frequencies of a test particle in the background spacetime. The twin higher frequencies are regarded respectively as the azimuthal frequency $\nu_{\phi}$ and the periastron precession frequency $\nu_{\text{per}}$ of a test-particle moving in quasi-circular orbits at the innermost disk region. The low-frequency mode in quasi-periodic oscillations is identified with the nodal precession frequency $\nu_{\text{nod}}$, which is emitted at the same radius where the twin higher frequencies signals are generated. With the observation data of GRO J1655-40 \cite{RPM1}, the constraint on the black hole parameters by quasi-periodic oscillations have been investigated in various theories of gravity \cite{TB1,TB101,TB2,TB3,TB4,TB5,TB6,TB7,TB8,TB9,TB10,TB11}. The main purpose of this paper is to constrain the parameters of a disformal Kerr black hole in quadratic DHOST theory by using of quasi-periodic oscillations in relativistic precession model together with the observation data of GRO J1655-40 \cite{RPM1}.

The paper is organized as follows: In Sec.II, we will review briefly the disformal Kerr black hole in quadratic DHOST \cite{dhostv1,dhostv2}. In Sec.III, we study quasi-periodic oscillations in the above disformal spacetime and then
make a constraint on the disformal parameter  with the observation data of GRO J1655-40. Finally, we present a summary.

\section{A disformal Kerr black hole in quadratic degenerate higher-order scalar-tensor theories}

In this section we review briefly a disformal Kerr black hole with an extra disformal parameter \cite{dhostv1,dhostv2}, which belongs to the non-stealth rotating solutions in quadratic DHOST theory. For the quadratic DHOST theory, the most general action can be expressed as \cite{dhost1}
\begin{eqnarray}
S=\int d^{4} x \sqrt{-g}\left(P(X, \varphi)+Q(X, \varphi) \square \varphi+F(X, \varphi) R+\sum_{i=1}^{5} A_{i}(X, \varphi) L_{i}\right),\label{S}
\end{eqnarray}
with
\begin{eqnarray}
L_{1} &\equiv& \varphi_{\mu \nu} \varphi^{\mu \nu}, \quad L_{2} \equiv(\square \varphi)^{2}, \quad L_{3} \equiv \varphi^{\mu} \varphi_{\mu \nu} \varphi^{\nu} \square \varphi, \nonumber\\
L_{4} &\equiv& \varphi^{\mu} \varphi_{\mu \nu} \varphi^{\nu \rho} \varphi_{\rho}, \quad L_{5} \equiv\left(\varphi^{\mu} \varphi_{\mu \nu} \varphi^{\nu}\right)^{2}.
\end{eqnarray}
Here  $R$ is the usual Ricci scalar and $\varphi$ is the scalar field with kinetic term  $X \equiv \varphi_{\mu} \varphi^{\mu}$, where the quantity $\varphi_{\mu}\equiv \nabla_{\mu} \varphi$ is covariant derivative of the scalar field. The functions $P$, $Q$, $F$ and  $A_i$ are only related to $\varphi$ and $X$. In quadratic DHOST theory, the functions $F$, $A_i$ must satisfy the so-called degeneracy conditions, but $P$ and $Q$ are  totally free, which ensures that there is only an extra scalar degree of freedom besides the usual tensor modes of gravity.  The degeneracy conditions for quadratic DHOST Ia theory is given in Refs.\cite{dhost3,dhost4}. The previous discussion tells us that a new solution  from  a ``seed" known solution  in DHOST Ia theory can be obtained by performing a disformal transformation of the metric. With the disformal transformation, the ``disformed" metric $g_{\mu \nu}$ is related to the original ``seed" metric $\tilde{g}_{\mu \nu}$  by \cite{dhost3}
\begin{eqnarray}
g_{\mu \nu}=A(X, \varphi) \tilde{g}_{\mu \nu}-B(X, \varphi) \varphi_{\mu} \varphi_{\nu},\label{metric1}
\end{eqnarray}
where $A(X,\varphi)$ and $B(X,\varphi)$ are  conformal and disformal factors, respectively.
In order to obtain a new solution,  the functions $A$ and $B$ must meet the conditions that the metrics $g_{\mu \nu}$ and $\tilde{g}_{\mu \nu}$ are not degenerate. Starting from the usual Kerr metric in general relativity, and setting the transformation functions $A(X, \varphi)=1$ and $B(X, \varphi)=B_0$ ($B_0$ is a constant), one can obtain the disformal Kerr metric  \cite{dhostv1,dhostv2}
\begin{eqnarray}
ds^{2}&=&-\frac{\Delta}{\rho}(d t-a\sin^{2}\theta d\phi)^{2}+\frac{\rho}{\Delta} dr^{2}+\rho d\theta^{2}+\frac{\sin ^{2}\theta}{\rho}(adt-(r^{2}+a^{2})d\phi)^{2} \nonumber\\
&&+\alpha(dt+\sqrt{2Mr(r^{2}+a^{2})}/\Delta dr)^{2},
\label{metric}
\end{eqnarray}
with
\begin{eqnarray}
\Delta=r^{2}+a^{2}-2 M r,\quad\quad\quad\rho=r^2+a^2\cos{\theta}^2.
\end{eqnarray}
Here the scalar field is taken as only a function of the coordinates $t$ and $r$ \cite{dhostv1,dhostv2}, i.e.,
\begin{eqnarray}
&&\phi(t, r)=-m t+S_{r}(r), \quad\quad\quad S_{r}= -\int  \frac{\sqrt{\mathcal{R}}}{\Delta} d r, \nonumber\\
&&\mathcal{R}=2 M m^{2} r(r^{2}+a^{2}),\quad\quad\quad \Delta=r^{2}+a^{2}-2 M r.\label{scalar}
\end{eqnarray}
The above form of scalar field can avoid the pathological behavior of the disformal metric at spatial infinite \cite{dhostv1,dhostv2}.
The parameter $\alpha$ is related to the rest mass $m$ of the scalar field by $\alpha=-B_{0}m^2$.
The choice of $A(X, \varphi)=1$ can avoid a global physically irrelevant constant conformal factor in the metric.
The negative sign in $S_r$ was chosen because the scalar field should be regular at the horizons of a Kerr black hole $\Delta=0$ \cite{dhostv1}. Comparing with the usual Kerr metric, one can find that the disformal Kerr metric (\ref{metric}) possesses a new disformal parameter $\alpha$ in quadratic DHOST theory, which encodes the deviation from general relativity. The presence of $\alpha$ in the metric (\ref{metric}) means that the disformal Kerr metric (\ref{metric}) is a non-stealth solution in quadratic DHOST theory because the scalar field $\varphi$ changes geometry of spacetime. As in the Kerr black hole case, the disformal Kerr metric (\ref{metric}) has also an intrinsic ring singularity at $\rho=0$ and the spacetime is asymptotically flat. However, it must be pointed out that its asymptotical behavior is not entirely the same as that of the Kerr one because the disformal Kerr metric (\ref{metric}) is not Ricci flat, i.e., $R_{\mu\nu}\neq0$. Moreover, the presence of the $drdt$ term yields the lack of circularity  in the disformal Kerr spacetime \cite{dhostv1,dhostv2}. In other words,  the spacetime can not be foliated by 2-dimensional meridional surfaces everywhere orthogonal to the Killing field $\xi=\partial_t$ and $\eta=\partial_{\phi}$ \cite{circle1,circle2,circle3}, which is qualitatively different from that in the usual Kerr spacetime in general relativity where spacetime is circular. The absence of circularity changes the structure of the black hole horizons so that the horizons depend on the polar angle $\theta$ and cannot be given by $r=const$ in Boyer-Lindquist coordinates, and then the corresponding surface gravity is no longer a constant \cite{dhostv1,dhostv2}.

\section{Constraint on parameters of a disformal Kerr black hole in quadratic degenerate
higher-order scalar-tensor theories by quasi-periodic oscillations}

In this section, we will make a constraint on parameters of the disformal Kerr black hole in quadratic DHOST by quasi-periodic oscillations. For a general stationary and axially symmetric spacetime with an additional $drdt$ term, its metric can be expressed as
\begin{eqnarray}
ds^2&=&g_{tt}dt^2+2g_{tr}dtdr+g_{rr}dr^2+2g_{t\phi}dtd\phi+g_{\theta\theta}d\theta^2
+g_{\phi\phi}
d\phi^2. \label{metric3n}
\end{eqnarray}
Since
the metric coefficients are independent of the coordinates $ t$ and $\phi$. one can find that for the geodesic motion of particle there exist two conserved quantities: the specific energy at infinity $E$ and the conserved $z$-component
of the specific angular momentum at infinity $L_z$. Due to the existence of an extra $drdt$ term, the forms of $E$ and $L_z$ are modified as
\begin{eqnarray}
E=-p_{t}=-g_{tt}\dot{t}-g_{tr}\dot{r}-g_{t\phi}\dot{\phi}, \quad \quad \quad L_{z}=p_{\phi}=g_{t\phi}\dot{t}+g_{\phi\phi}\dot{\phi}.\label{conserved quantities}
\end{eqnarray}
And then the timelike geodesics can be expressed as
\begin{eqnarray}
&&\dot{t}=\frac{g_{\phi\phi}E+g_{t\phi}L_z+g_{tr}g_{\phi\phi}\dot{r}}{g_{t\phi}^2-g_{tt}g_{\phi\phi}},\label{u1}\\
&&\dot{\phi}=\frac{g_{t\phi}E+g_{tt}L_z+g_{tr}g_{t\phi}\dot{r}}{g_{tt}g_{\phi\phi}-g_{t\phi}^2},\label{u2}\\
&&\tilde{g}_{rr}\dot{r}^2+g_{\theta\theta}\dot{\theta}^2=V_{eff}(r,\theta; E,L_z),\label{u3}
\end{eqnarray}
with the effective potential
\begin{eqnarray}
\tilde{g}_{rr}=\bigg[g_{rr}+\frac{g_{tr}^2g_{\phi\phi}}{g_{t\phi}^2-g_{tt}g_{\phi\phi}}\bigg],\quad\quad\quad
V_{eff}(r,\theta; E,L_z)=\frac{E^2g_{\phi\phi}+2EL_zg_{t\phi}+L^2_zg_{tt}
}{g^2_{t\phi}-g_{tt}g_{\phi\phi}}-1,
\end{eqnarray}
where the overhead dot represents a derivative with respect to the
affine parameter $\lambda$. The radial component of geodesic equations
\begin{eqnarray}
\frac{d}{d\lambda}(g_{\mu\nu}\dot{x}^{\nu})=\frac{1}{2}(\partial_{\mu}g_{\nu\rho})\dot{x}^{\nu}\dot{x}^{\rho},
\end{eqnarray}
can be written as \cite{RPM1,RPM2,RPM20,RPM3}
\begin{eqnarray}\label{cedx0r}
\frac{d}{d\lambda}(g_{rr}\dot{r}+g_{tr}\dot{t}\; )=\frac{1}{2}\bigg[(\partial_{r}g_{tt})\dot{t}^2+2(\partial_{r}g_{t\phi})\dot{t}\dot{\phi}+(\partial_{r}g_{\phi\phi})\dot{\phi}^2+
2\partial_{r}g_{tr})\dot{t}\dot{r}+\partial_{r}g_{rr})\dot{r}^2+\partial_{r}g_{\theta\theta})\dot{\theta}^2\bigg].
\end{eqnarray}
We here consider only the case where a particle moving along a circular orbit in the equatorial plane, i.e., $r=r_0$ and $\theta=\pi/2$, which means that $\dot{r}=\dot{\theta}=\ddot{r}=0$. From Eq.(\ref{u1}), we can find that $\dot{t}$ can be rewritten as a form
\begin{eqnarray}
\dot{t}=P(r,\theta)+Q(r,\theta)\dot{r},
\end{eqnarray}
with
\begin{eqnarray}
P(r,\theta)\equiv\frac{g_{\phi\phi}E+g_{t\phi}L_z}{g_{t\phi}^2-g_{tt}g_{\phi\phi}},\quad\quad\quad
Q(r,\theta)\dot{r}\equiv\frac{g_{tr}g_{\phi\phi}}{g_{t\phi}^2-g_{tt}g_{\phi\phi}}.
\end{eqnarray}
And then $\ddot{t}$ can be further expressed as
\begin{eqnarray}
\ddot{t}=[\partial_{r}P(r,\theta)]\dot{r}+[\partial_{\theta}P(r,\theta)]\dot{\theta}+[\partial_{r}Q(r,\theta)]\dot{r}^2+
[\partial_{\theta}Q(r,\theta)]\dot{r}\dot{\theta}+Q(r,\theta)\ddot{r}.
\end{eqnarray}
Thus, for the circular equatorial orbit case we considered, one has $\ddot{t}=0$, and then the left side of Eq.(\ref{cedx0r}) becomes
\begin{eqnarray}
\frac{d}{d\lambda}(g_{rr}\dot{r}+g_{tr}\dot{t}\; )=(\partial_{r}g_{rr})\dot{r}^2+(\partial_{\theta}g_{rr})\dot{r}\dot{\theta}+
(\partial_{r}g_{tr})\dot{t}\dot{r}+(\partial_{\theta}g_{tr})\dot{t}\dot{\theta}+g_{tr}\ddot{t}=0,
\end{eqnarray}
 which means that in this case Eq.(\ref{cedx0r}) can be simplified as
\begin{eqnarray}\label{tdot0}
(\partial_{r}g_{tt})\dot{t}^2+2(\partial_{r}g_{t\phi})\dot{t}\dot{\phi}+(\partial_{r}g_{\phi\phi})\dot{\phi}^2=0.
\end{eqnarray}
Thus, the orbital angular velocity $\Omega_{\phi}$ of particle moving in the circular orbits has a form
\begin{eqnarray}
\Omega_{\phi}=\frac{d\phi}{dt}=\frac{-g_{t\phi,r}\pm\sqrt{(g_{t\phi,r})^2
+g_{tt,r}g_{\phi\phi,r}}}{g_{\phi\phi,r}},\label{jsd0}
\end{eqnarray}
where the sign is $+ (-)$ for corotating (counterrotating)
orbits. The corresponding azimuthal frequency $\nu_{\phi}=\Omega_{\phi}/(2\pi)$.
On the other hand, for a timelike particle moving along circular orbits in the equatorial plane, it must satisfy the conditions
\begin{eqnarray}
g_{\mu\nu}\dot{x}^{\mu}\dot{x}^{\nu}=-1,\quad\quad\quad \dot{r}=\dot{\theta}=0.
\end{eqnarray}
which leads to another relationship between $\dot{t}$ and $\dot{\phi}$
\begin{eqnarray}\label{tdot1}
g_{tt}\dot{t}^2+2g_{t\phi}\dot{t}\dot{\phi}+g_{\phi\phi}\dot{\phi}^2=-1.
\end{eqnarray}
Two equations (\ref{tdot0}) and (\ref{tdot1}) are  independent of each other because the former is obtained by geodesic equations and the latter is gotten by the timelike condition of particle. Making use of these two equations, one can find
\begin{eqnarray}\label{t0}
\dot{t}=\frac{1}{\sqrt{-g_{tt}-2g_{t\phi}\Omega_{\phi}-g_{\phi\phi}\Omega_{\phi}^2}}.
\end{eqnarray}
Combining Eq.(\ref{conserved quantities}) with the condition $\dot{r}=0$, one can obtain \cite{RPM1,RPM2,RPM20,RPM3}
\begin{eqnarray}
&&E=-\frac{g_{tt}+g_{t\phi}\Omega_{\phi}}{\sqrt{-g_{tt}-2g_{t\phi}\Omega_{\phi}
-g_{\phi\phi}\Omega^2_{\phi}}},\nonumber\\
&&L_z=\frac{g_{t\phi}+g_{\phi\phi}\Omega_{\phi}}{\sqrt{-g_{tt}
-2g_{t\phi}\Omega_{\phi}-g_{\phi\phi}\Omega^2_{\phi}}}.\label{jsd}
\end{eqnarray}
 Considering a small perturbation around a circular equatorial orbit \cite{TB1,TB101,TB2,TB3,TB4,TB5,TB6,TB7,TB8,TB9,TB10,TB11}, i.e.,
\begin{eqnarray}
r(t)=r_0+\delta r(t), \;\;\;\;\;\;\;\;\;\;\theta(t)=\frac{\pi}{2}+\delta \theta(t).
\end{eqnarray}
one can find that the perturbations $\delta r(t)$ and $\delta \theta(t)$ are governed by the following differential equations
\begin{eqnarray}
\frac{d^2\delta r(t) }{dt^2}+\Omega^2_{r}\delta r(t)=0, \;\;\;\;\;\;\;\;\;\;\frac{d^2\delta \theta(t) }{dt^2}+\Omega^2_{\theta}\;\delta \theta(t)=0,
\end{eqnarray}
with
\begin{eqnarray}
\Omega^2_{r}=-\frac{1}{2\tilde{g}_{rr}\dot{t}^2}\frac{\partial^2 V_{eff}}{\partial r^2}\bigg|_{r=r_0,\theta=\frac{\pi}{2}}, \;\;\;\;\;\;\;\;\;\;\Omega^2_{\theta}=-\frac{1}{2g_{\theta\theta}\dot{t}^2}\frac{\partial^2 V_{eff}}{\partial \theta^2}\bigg|_{r=r_0,\theta=\frac{\pi}{2}}.\label{jsdd0}
\end{eqnarray}
The radial epicyclic frequency $\nu_r$ and the vertical
epicyclic frequency $\nu_{\theta}$ can be written as $\nu_r=\Omega_r/2\pi$ and $\nu_{\theta}=\Omega_{\theta}/2\pi$, respectively. Inserting metric functions (\ref{metric}) into Eq.(\ref{jsd0}) ,
we can find the azimuthal frequency
\begin{eqnarray}\label{pinlv1}
\nu_{\phi}=\frac{1}{2\pi}\frac{M^{1/2}}{r^{3/2}+a^{*}M^{3/2}},
\end{eqnarray}
where $a^{*}\equiv a/M$. Obviously, the frequency $\nu_{\phi}$ is independent of the parameter $\alpha$ and owns the same form as that in the usual Kerr black hole spacetime \cite{RPM1,RPM2,RPM20,RPM3}. It can be explained by a fact that $g_{t\phi}$ and $g_{\phi\phi}$ in the metric (\ref{metric}) do not contain $\alpha$, while in  $g_{tt}$, the term related to $\alpha$ is independent of the radial coordinate $r$ so that the parameter $\alpha$ vanishes in the term $g_{tt,r}$ and in the orbital angular velocity $\Omega_{\phi}$. Similarly, substituting metric functions (\ref{metric}) into Eqs.(\ref{t0}) and (\ref{jsdd0}), one has
\begin{eqnarray}
&&\dot{t}^2=\frac{(r^{3/2}+a\sqrt{M})^2}{W},\quad\quad\quad\frac{1}{g_{\theta\theta}}\frac{\partial^2 V_{eff}}{\partial \theta^2}=-\frac{2M(r^2+3a^2-4a\sqrt{M r})}{r^2W},\nonumber\\
&&\frac{1}{\tilde{g}_{rr}}\frac{\partial^2 V_{eff}}{\partial r^2}=-\frac{2M[2M(3r^2-\alpha a^2)-(1-\alpha)r(r^2-3a^2+8a\sqrt{M r})]}{(1-\alpha)r^3W},\\
&&W\equiv M(3r^2+\alpha a^2)-(1-\alpha)r^{3/2}(r^{3/2}+2a\sqrt{M})\nonumber.
\end{eqnarray}
We note that the denominators in above three quantities contain the same factor $W$, which can be exactly canceled out in the frequencies $\nu_{r}$ and  $\nu_{\theta}$. This leads to that the vertical
epicyclic frequency $\nu_{\theta}$ is not a function of $\alpha$ and the radial epicyclic frequency $\nu_{r}$ has a simple form, i.e.,
\begin{eqnarray}
\nu_{r}&=&\nu_{\phi}\bigg[1-\frac{6M}{(1-\alpha)r}+8a^{*}\frac{M^{3/2}}{r^{3/2}}-3a^{*2}\frac{M^{2}}{r^{2}}
+\frac{2\alpha a^{*2}}{1-\alpha}\frac{M^{3}}{r^{3}}\bigg]^{1/2},\label{pinlv2}\\
\nu_{\theta}&=&\nu_{\phi}\bigg[1-4a^{*}\frac{M^{3/2}}{r^{3/2}}+3a^{*2}\frac{M^{2}}{r^{2}}\bigg]^{1/2}.\label{pinlv22}
\end{eqnarray}
The  dependence of  $\nu_r$ on the parameter $\alpha$ make it possible to constrain the disformal effect by quasi-periodic oscillations.
Finally, the periastron and nodal precession frequencies can be expressed as
\begin{eqnarray}
\nu_{\text{per}}=\nu_{\phi}-\nu_{r},\;\;\;\;\;\;\;\;\;\;\;
\nu_{\text{nod}}=\nu_{\phi}-\nu_{\theta},
\end{eqnarray}
respectively.
\begin{figure}[ht]
\begin{center}
\includegraphics[width=6cm]{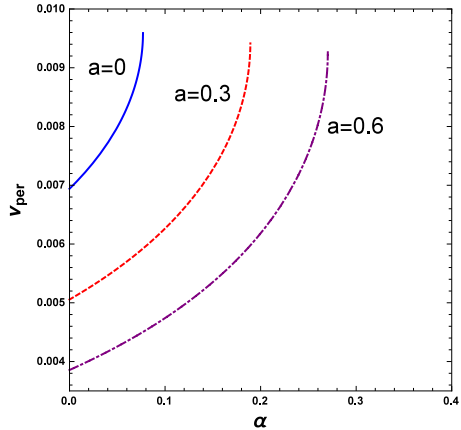}
\caption{The change of the frequency $\nu_{\text{per}}$  with the parameter $\alpha$ in a disformal Kerr black hole spacetime in DHOST theories. Here we set $M=1$ and $r=6.5$.}\label{fig1}
\end{center}
\end{figure}
From Eqs. (\ref{pinlv1}),  it is obvious that the nodal precession frequency $\nu_{\text{nod}}$ is independent of the parameter $\alpha$. Thus, in Fig.(\ref{fig1}), we plot only the change of $\nu_{\text{per}}$ for the disformal Kerr black hole spacetime. It is shown that the periastron precession frequency $\nu_{\text{per}}$ increases with the parameter $\alpha$, but decreases with the spin parameter $a$.

From previous analysis, one can find that there are three simultaneous quasi-periodic oscillations frequencies which are generated at the same radius of the orbit in the accretion flow. For a Kerr black hole spacetime in general relativity,  one can  determine the three variables ($r$, $M$, and $a$) with observed data of three quasi-periodic frequencies by solving directly the corresponding three equations. However, for a disformal Kerr black hole spacetime in DHOST theories, there is an extra disformal parameter $\alpha$ which describes its deviation from usual Kerr black hole. Thus, we have to resort to the $\chi^2$ analysis and best-fit the values of four unknown variables in the disformal Kerr black hole spacetime in DHOST theories (\ref{metric}).
From the current observations of GRO J1655-40, one can obtain two set of data about these frequencies ($\nu_{\phi}, \nu_{\text{per}},\nu_{\text{nod}}$ )\cite{RPM1,TB1}:
\begin{eqnarray}
(441^{+2}_{-2},\;\;298^{+4}_{-4}, \;\;17.3^{+0.1}_{-0.1})\;\;\;\; \text{and}\;\;\;\;
(451^{+5}_{-5},\;\;-, \;\;18.3^{+0.1}_{-0.1}).
\end{eqnarray}
Moreover, there also is an independent dynamical measurement
of the mass of the black hole \cite{TB4}: $M_{\text{dyn}}=5.4\pm0.3M_{\odot}$. Therefore, there are five free parameters: mass $M$, spin parameter $a$, the disformal parameter $\alpha$,
the radius $r_1$ and $r_2$ correspond the observations with three frequencies and two frequencies, respectively.
With these data, we can test the gravity of a disformal Kerr black hole (\ref{metric}) through the relativistic precession model as in ref.\cite{TB1}. Performing a $\chi^2$ analysis,  we obtain the minimum $\chi^2_{\text{min}}=0.1946$ and constrain the black hole parameters
\begin{eqnarray}
M=5.400^{+0.038}_{-0.037}M_{\odot},\;\;\;\;\;\;\;\;a^{*}=0.281^{+0.004}_{-0.003},
\;\;\;\;\;\;\;\;\alpha=-0.010^{+0.011}_{-0.012},
\end{eqnarray}
at the $68.3\%$ C.L.
\begin{figure}[ht]
\begin{center}
\includegraphics[width=5cm]{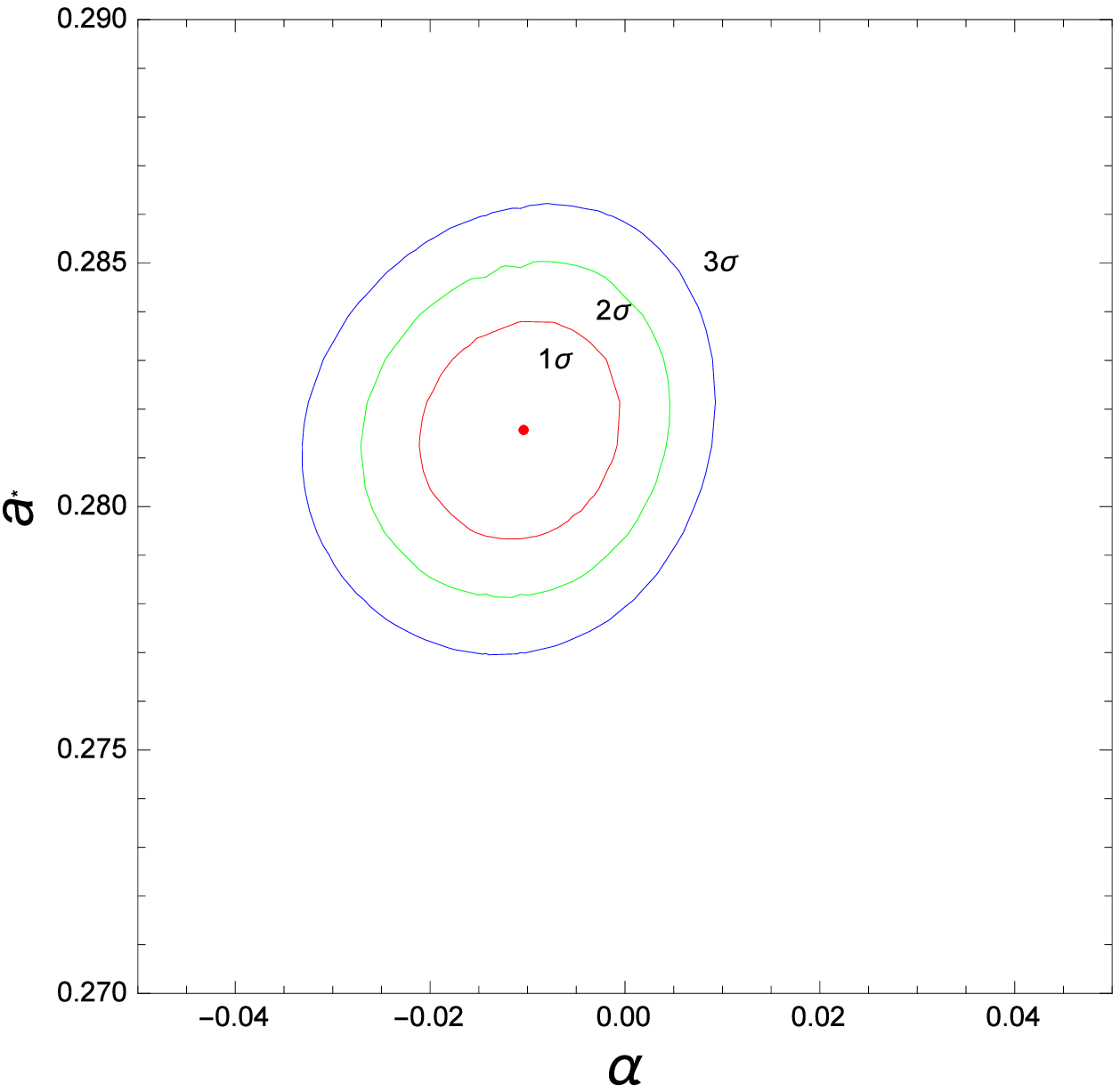}\;\;\;\;\includegraphics[width=5cm]{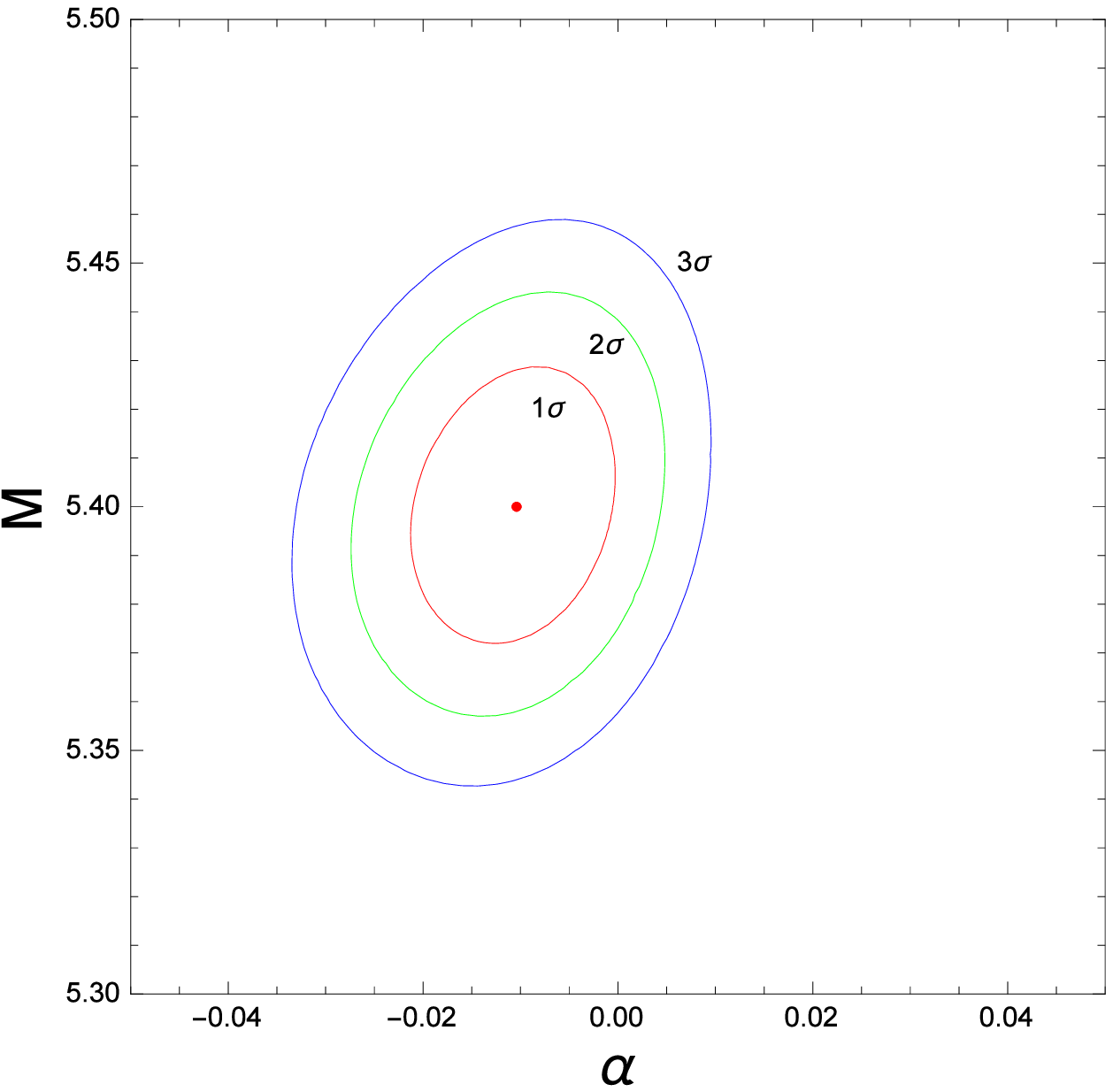}\;\;\;\;\includegraphics[width=5.1cm]{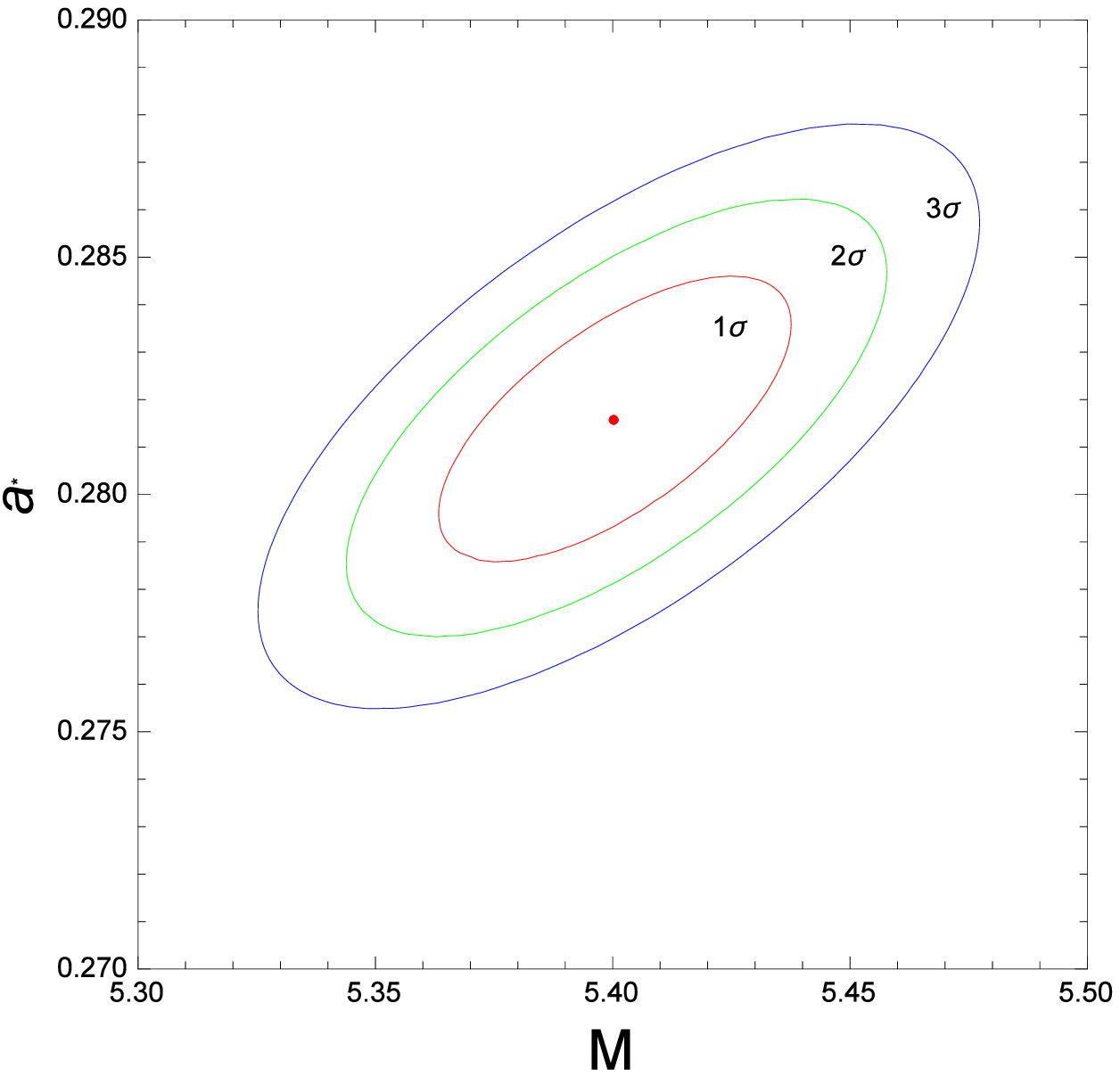}
\caption{Constraints on the parameters of a disformal Kerr black hole (\ref{metric}) with GRO J1655-40 from current observations of QPOs within the relativistic precession model. The red, green and blue lines represent the contour levels $1\sigma$, $2\sigma$ and $3\sigma$, respectively. The red dot in the panels correspond the best-fit values of parameters: $M=5.400$, $a^{*}=0.281$ and $\alpha=-0.010$.}\label{fig2}
\end{center}
\end{figure}
 The best-fit values of the radius of circular orbital corresponding two sets of quasi-periodic oscillations are $r_1=5.619 M=1.1292\;r_{\text{ISCO}}$ and $r_2=5.515 M=1.1083 \;r_{\text{ISCO}}$, respectively. Here $r_{\text{ISCO}}$ is the innermost stable circular orbit in the disformal Kerr black hole  spacetime (\ref{metric}) with the best-fit values ($M=5.400$, $a^{*}=0.281$ and $\alpha=-0.010$), which means that the circular orbit of quasi-periodic oscillations lies in the strong gravitational-field region of the black hole. In Fig.(\ref{fig2}), we show the contour levels of $1\sigma$, $2\sigma$ and $3\sigma$ for the black hole parameters $M$, $a$ and $\alpha$. Our results show that the disformal parameter $\alpha$ is almost negative in the range of $1\sigma$, which implies the negative disformal parameter $\alpha$ is favored by the observation data of GRO J1655-40. The non-zero best-fit value of $\alpha$ indicates that the spacetime described gravitational field of the source of GRO J1655-40 should be not Ricci flat and  the surface gravity of horizon is not a constant due to the absence of circularity \cite{dhostv1,dhostv2}. Moreover, the negative best-fit value of $\alpha$ means that the outer ergosurface obeyed $r_{outerg}=M/(1-\alpha)+\sqrt{M^2/(1-\alpha)^2-a^2\cos^2\theta}$ is smaller than that in the usual Kerr spacetime for fixed $\theta$, which will modify the possibility of exacting energy by Penrose process.  The negative  $\alpha$ also yields that a null surface defined by Eq.(23) in ref.\cite{dhostv2}, a candidate event horizon, has a minimum at the equator. Meanwhile, it implies that the width of ergosphere owns a maximum at the equator.

\section{summary}

With relativistic precession model, we have studied  quasi-periodic oscillations frequencies in a disformal Kerr black hole in quadratic degenerate higher-order scalar-tensor theories. The black hole owns three parameters: mass $M$, spin $a$ and the disformal parameter $\alpha$. We find that the periastron precession frequency $\nu_{\text{per}}$ increases with $\alpha$ for different spin parameters, while the azimuthal frequency $\nu_{\phi}$ and the nodal precession frequency $\nu_{\text{nod}}$ are independent of the parameter $\alpha$. With increase of the spin parameter $a$ of black hole, $\nu_{\phi}$ and $\nu_{\text{per}}$ decrease, but $\nu_{\text{nod}}$ increases, which are similar to those in the usual rotating black hole spacetimes.
With the observation data of GRO J1655-40, we constrain the parameters of the disformal Kerr black hole in quadratic degenerate higher-order scalar-tensor theories, and find that in the range of $1 \sigma$ the disformal parameter $\alpha$ is almost negative. This implies the negative disformal parameter $\alpha$ is favored by the observation data of GRO J1655-40.

\section{\bf Acknowledgments}
This work was  supported by the National Natural Science
Foundation of China under Grant No.11875026, 11875025, 12035005 and 2020YFC2201403.


\vspace*{0.2cm}
\begin{thebibliography}{99}

\baselineskip=0.6 cm \baselineskip=0.6 cm

\bibitem{P1} B. P. Abbott et al., \textit{Observation of Gravitational Waves from a Binary Black Hole Merger},  Phys. Rev. Lett. {\bf116}, 061102 (2016), [arXiv:1602.03837].
\bibitem{P2} B. P. Abbott et al., \textit{GW151226: Observation of Gravitational Waves from a 22-Solar-Mass Binary Black Hole Coalescence}, Phys. Rev. Lett. {\bf116}, 241103 (2016), [arXiv:1606.04855].
\bibitem{P3} B. P. Abbott et al.,\textit{GW170104: Observation of a 50-Solar-Mass Binary Black Hole Coalescence at Redshift 0.2} Phys. Rev. Lett. {\bf118}, 221101 (2017), [arXiv:1706.01812].
\bibitem{P4} B. P. Abbott et al., \textit{GW170814: A Three-Detector Observation of Gravitational Waves from a Binary Black Hole Coalescence}, Phys. Rev. Lett. {\bf198}, 141101 (2017), [arXiv:1709.09660].
\bibitem{P5} B. P. Abbott et al., \textit{GW170608: Observation of a 19-solar-mass Binary Black Hole Coalescence}, Astrophys. J. {\bf851}, L35 (2017), [arXiv:1711.05578].

\bibitem{fbhs1} The Event Horizon Telescope Collaboration, \textit{First M87 Event Horizon Telescope Results. I. The Shadow of the Supermassive Black Hole}, Astrophys. J. Lett. {\bf875}, L1 (2019).
\bibitem{fbhs6} The Event Horizon Telescope Collaboration, \textit{First M87 Event Horizon Telescope Results. VI. The Shadow and Mass of the Central Black Hole}, Astrophys. J. Lett. {\bf875}, L6 (2019).



\bibitem{dhost1} D. Langlois and K. Noui, \textit{Degenerate higher derivative theories beyond Horndeski: evading the
Ostrogradski instability}, J. Cosmol. Astropart. Phys. {\bf02},034 (2016).
\bibitem{dhost2} M. Crisostomi, K. Koyama, G. Tasinato, \textit{Extended Scalar-Tensor Theories of Gravity}, J. Cosmol. Astropart. Phys.{\bf04}, 044 (2016).
\bibitem{dhost3} D. Langlois and K. Noui, \textit{Hamiltonian analysis of higher derivative scalar-tensor theories}, J. Cosmol. Astropart. Phys. {\bf07},016 (2016).

\bibitem{dhost4} J. Achour, D. Langlois, and K. Noui, \textit{Degenerate higher order scalar-tensor theories beyond
Horndeski and disformal transformations}, Phys. Rev. D {\bf93},124005 (2016),  [arXiv:1602.08398].

\bibitem{late2} G. W. Horndeski, \textit{Second-order scalar-tensor field equations in a four-dimensional space}, Int. J. Theor. Phys. {\bf 10} 363-384 (1974).
\bibitem{late3} J. Gleyzes, D. Langlois, F. Piazza, and F. Vernizzi, \textit{Healthy theories beyond Horndeski}, Phys. Rev. Lett. { \bf114}, 211101  (2015).



\bibitem{c2633} J. Achour and H. Liu, \textit{Hairy Schwarzschild-(A)dS black hole solutions in DHOST theories beyond shift symmetry}, Phys. Rev. D {\bf99}, 064042 (2019), [arXiv:1811.05369].
\bibitem{c2634} H. Motohashi and M. Minamitsuji, \textit{Exact black hole solutions in shift-symmetric quadratic degenerate higher-order scalar-tensor theories}, Phys. Rev. D {\bf99}, 064040 (2019), [arXiv:1901.04658].
\bibitem{c2635} M. Minamitsuji and J. Edholm, \textit{Black hole solutions in shift-symmetric degenerate higher-order scalar-tensor theories}, Phys. Rev. D {\bf100}, 044053 (2019), [arXiv:1907.02072].
\bibitem{dhost24} C. Charmousis, M. Crisostomi, R. Gregory, and N. Stergioulas, \textit{Rotating Black Holes in Higher
Order Gravity}, Phys. Rev. D {\bf100} 084020, (2019), [arXiv:1903.05519].
\bibitem{dhost26} K. Takahashi and H. Motohashi, \textit{General Relativity solutions with stealth scalar hair in quadratic
higher-order scalar-tensor theories},  J. Cosmol. Astropart. Phys. {\bf06}, 034 (2020), [arXiv:2004.03883].
\bibitem{dhost20} J. Achour, H. Liu, and S. Mukohyama, \textit{Hairy black holes in DHOST theories: Exploring
disformal transformation as a solution-generating method}, J. Cosmol. Astropart. Phys. {\bf02}, 023 (2020), [arXiv:1910.11017].

\bibitem{dhost27} E. Babichev, C. Charmousis, A. Cisterna, and M. Hassaine, \textit{Regular black holes via the Kerr-Schild
construction in DHOST theories}, J. Cosmol. Astropart. Phys. {\bf06}, 049 (2020),[arXiv:2004.00597].
\bibitem{dhost28} K. V. Aelst, E. Gourgoulhon, P. Grandcl\'{e}ment, and C. Charmousis, \textit{Hairy rotating black holes in
cubic Galileon theory}, Class. Quant. Grav. {\bf37}, 035007 (2020), [arXiv:1910.08451].



\bibitem{dhostv2} J.  Achour, H. Liu, H. Motohashi, S. Mukohyama, and K. Noui, \textit{On Rotating Black Holes in DHOST Theories}, J. Cosmol. Astropart. Phys. {\bf11}, 001 (2020), [arXiv:2006.07245].
\bibitem{dhostv1} T. Anson, E. Babichev, C. Charmousis, and M. Hassaine, \textit{Disforming the Kerr metric}, J. High Energ. Phys. {\bf01}, 018 (2021), [arXiv:2006.06461].
\bibitem{shadow1} F. Long, S. Chen, M. Wang, J. Jing, \textit{Shadow of a disformal Kerr black hole in quadratic degenerate
higher-order scalar-tensor theories}, Eur. Phys. J. C {\bf80}, 1180 (2020).



\bibitem{qpo1} R. A. Remillard and J. E. McClintock, \textit{X-ray Properties of Black-Hole Binaries}, Ann. Rev. Astron.
Astrophys. {\bf44}, 49 (2006).
\bibitem{qpo2} T. M. Belloni and S. E. Motta, \textit{Transient Black Hole Binaries}, arXiv:1603.07872.


\bibitem{RPM1} S. E. Motta, T. M. Belloni, L. Stella, T. Muoz-Darias
and R. Fender, \textit{Precise mass and spin measurements for a stellar-mass black hole through X-ray timing: the case of GRO J1655-40}, Mon. Not. Roy. Astron. Soc. {\bf437}, 2554 (2014) [arXiv:1309.3652 [astro-ph.HE]].

\bibitem{RPM2} S. E. Motta, T. Muoz-Darias, A. Sanna, R. Fender,
T. Belloni and L. Stella, \textit{Black hole spin measurements through the relativistic precession model: XTE J1550-564}, Mon. Not. Roy. Astron. Soc. {\bf439}, 65 (2014) [arXiv:1312.3114 [astro-ph.HE]].
\bibitem{RPM20} P. Casella,  T. Belloni, and L. Stella, \textit{The ABC of low-frequency quasi-periodic oscillations in black-hole candidates: Analogies with Z-sources}, Astrophys. J. {\bf629}, 403 (2005).

\bibitem{RPM3} L. Stella and M. Vietri, \textit{Lense-Thirring Precession and QPOs in Low Mass X-Ray Binaries }, Astrophys. J. {\bf492}, L59 (1998) [astro-ph/9709085]
\bibitem{RPM301} L. Stella and M. Vietri, \textit{kHz Quasi Periodic Oscillations in Low Mass X-ray Binaries as Probes of General Relativity in the Strong Field Regime }, Phys. Rev. Lett. {\bf82}, 17 (1999) [astro-ph/9812124].
\bibitem{RPM302} L. Stella, M. Vietri and S. Morsink, \textit{Correlations in the QPO Frequencies of Low Mass X-Ray Binaries and the Relativistic Precession Model }, Astrophys. J. {\bf524}, L63 (1999) [astro-ph/9907346].

\bibitem{TB1} C. Bambi, \textit{Probing the space-time geometry around black hole candidates with the resonance models for high-frequency QPOs and comparison with the continuum-fitting method}, J. Cosmol. Astropart. Phys. {\bf1209}, 014 (2012)

\bibitem{TB101}C. Bambi and S. Nampalliwar, \textit{Quasi-periodic oscillations as a tool for testing the Kerr metric: A comparison with gravitational waves and iron line},Europhys. Lett. {\bf116}, 30006 (2016), arXiv:1604.02643.
\bibitem{TB2} Z. Stuchlik and A. Kotrlova, \textit{Orbital resonances in discs around braneworld Kerr black holes}, Gen. Rel. Grav. {\bf41}, 1305 (2009).
\bibitem{TB3} T. Johannsen and D. Psaltis, \textit{Testing the No-Hair Theorem with Observations in the Electromagnetic Spectrum. III. Quasi-Periodic Variability}, Astrophys. J. {\bf726}, 11 (2011) [arXiv:1010.1000 [astro-ph.HE]].
\bibitem{TB4} M. E. Beer and P. Podsiadlowski, \textit{The quiescent light curve and evolutionary state of GRO J1655-40}, Mon. Not. Roy. Astron. Soc. {\bf331}, 351 (2002) [astro-ph/0109136].
\bibitem{TB5} A. Maselli, L. Gualtieri, P. Pani, L. Stella, and V. Ferrari, \textit{Testing Gravity with Quasi Periodic Oscillations from accreting Black Holes: the Case of Einstein-Dilaton-Gauss-Bonnet Theory}, Astrophys. J. {\bf801}, 115 (2015).

\bibitem{TB6}A. G. Suvorov and A. Melatos,  \textit{Testing modified gravity and no-hair relations for the Kerr-Newman metric through quasiperiodic oscillations of galactic microquasars}, Phys. Rev. D {\bf93}, 024004 (2016).
\bibitem{TB7}G. Pappas, \textit{What can quasi-periodic oscillations tell us about the structure of the corresponding compact objects?}, Mon. Not. R. Astron. Soc. {\bf422}, 2581-2589 (2012).

\bibitem{TB8} K. Boshkayev, D. Bini, J. Rueda, A. Geralico, M. Muccino and I. Siutsou, \textit{What can we extract from quasiperiodic oscillations?}, Grav. Cosmol. {\bf20}, 233-239 (2014).

\bibitem{TB9} S. Chen, M. Wang, J.Jing, \textit{Testing gravity of a regular and slowly rotating phantom black hole by quasiperiodic oscillations}, Class. Quantum Grav. {\bf33}, 195002 (2016).

 \bibitem{TB10} A. Allahyari,  L. Shao,   \textit{Testing No-Hair Theorem by Quasi-Periodic Oscillations: the quadrupole of GRO J1655-40}, arXiv: 2102.02232.
 \bibitem{TB11}A. Maselli, L. Gualtieri, P. Pani, L. Stella, V. Ferrari,  \textit{Testing Gravity with Quasi Periodic Oscillations from accreting Black Holes: the Case of Einstein-Dilaton-Gauss-Bonnet Theory},  Astrophys. J. {\bf801} 2, 115 (2015).

 \bibitem{circle1} V. Frolov and I. Novikov, \textit{Black Hole Physics, Basic Concepts and New Developments}, Fundam. Theor. Phys. {\bf96} (1998) doi:10.1007/978-94-011-5139-9.
\bibitem{circle2}  E. Gourgoulhon and S. Bonazzola,  \textit{Noncircular axisymmetric stationary spacetimes}, Phys. Rev. D {\bf48}, 2635 (1993).
\bibitem{circle3} K. V. Aelst, E. Gourgoulhon, P. Grandcl\'{e}ment, and C. Charmousis, \textit{Hairy rotating black holes in cubic Galileon theory}, Class. Quantum Grav. {\bf37}, 035007 (2020).

\end{thebibliography}
\end{document}